# Polyurethane-Based Scintillators for Neutron and Gamma Radiation Detection in Medical and Industrial Applications

Olga Maiatska[a], Torsten Dünnebacke[a], Martin Kreuels[a], Guntram Pausch[a*], Falko Scherwinski[a], and Jürgen R. Stein[a]

[a]*Rapiscan Systems GmbH, Heinz-Fangman-Str. 4, 42287 Wuppertal, Germany*

**Abstract**

Organic scintillators using a solid polyurethane (PU) matrix have been introduced to combine the robustness of a construction material with scintillating properties that allow gamma rays and fast neutrons to be detected efficiently and at low cost. This work compares two corresponding materials, the older M600 and the more recent M700, with EJ-276D and EJ-200 representing common plastic scintillators with and without pulse-shape discrimination (PSD) capabilities, respectively. Characterization measurements were performed with small samples of 26 mm diameter and 10 mm height, which were coupled to a photomultiplier tube (PMT) and simultaneously exposed to $^{252}$Cf fission neutrons and $^{137}$Cs gamma rays. M700 turned out to provide the best PSD performance and about the same light yield as EJ-276D, while its light pulses exhibit a shorter pulse decay. An accelerated ageing process applied in between two test campaigns was too short to trigger distinct performance degradation in any of the materials, though optical degradation was visible in EJ-276D and in EJ-200 but not in the PU-based materials. Nevertheless, the extremely robust polyurethane matrix promises advantages in medical and industrial applications where resilience and long-term stability are of crucial importance.

## 1. Introduction

Solid organic scintillators have been used to measure neutron and gamma radiation for more than half a century [1]-[2]. Their hydrogen content makes them sensitive to fast neutrons, as (elastic) neutron scattering on hydrogen nuclei – consisting of a single proton – generates fast recoil protons and corresponding ionization tracks within the scintillator, which are easily detected. Gamma detection in organic scintillators is basically due to Compton scattering, as low charge numbers of the scintillators' constituents make photoabsorption quite improbable. Consequently, energy spectra of gamma rays measured with plastic scintillators do not exhibit photopeaks but characteristic Compton edges, at least in the energy range of 30 keV to 3 MeV, which is the most interesting one for many applications. This and the moderate light yield of common plastic materials (around 10 photons / keV [3]), which cannot compete with that of inorganic crystals as NaI(Tl) or LaBr$_3$(Ce) (about 45 to 76 photons/keV, respectively [4]), result in a poor spectroscopic performance. However, plastic scintillators can be produced in large sizes and are quite affordable. Their unrivaled sensitivity per cost (counts per second and dollar of detector investment in a given radiation field) makes them attractive for many applications in homeland security, industry, or medicine.

---

* Corresponding author. e-mail: gpausch@rapiscan.com

The favorable sensitivity-to-cost ratio is the reason why large-volume plastic scintillators have been used in Radiation Portal Monitors (RPMs), built to disclose potentially illicit transportation of radioactive material through points of entrance (POE) by measuring gamma radiation emitted from the cargo. However, ageing effects in common polyvinyl toluene (PVT) scintillators – especially the so-called "fogging" by exposure to humidity in combination with large temperature variations – turned out to be a weak point limiting the lifetime of corresponding detectors [5]. This issue motivated Target Systemelektronik[1] some years ago to start research in plastic scintillators. The initial question was: Is it possible to provide extremely robust and rugged plastic materials as used in civil engineering or machine construction with scintillating properties? Corresponding efforts [6] lead to a first suitable polyurethane-(PU-)based scintillator material named M600, available since 2018, which exhibits not only reasonable light yield but also quite decent pulse-shape discrimination (PSD) capabilities [7]. The latter is an inestimable advantage, as it allows to discriminate neutron signals (recoil protons) against gamma detections (recoil electrons) and thus to find neutron signatures in a gamma background, which is important for a doubtless detection and identification of Special Nuclear Materials (SNM). Note that the most common solid organic scintillators including those used in RPMs cannot distinguish neutron- from gamma-induced events.

PSD-capable plastic scintillators had been developed [8] and marketed before PU-based materials were available. An early commercial product, the Eljen E-J276, could provide reasonable discrimination (see e.g. [9]) but suffered from ageing in terms of a significant reduction of light yield over time [10]. Meanwhile, several new materials have been developed and introduced [11]-[12], in particular organic glass scintillators (OGS) [13]-[14] which outperform other plastic scintillators in terms of light yield and PSD performance. This invention facilitated, amongst others, new projects of dual-particle (neutron and gamma-ray) imagers in homeland security (see e.g. [15]) and medicine [16]-[17] using arrangements of multiple plastic scintillator bars. The overall performance and imaging quality in such projects depend, among other things, on the scintillators' light yield (affecting energy, time, and depth-of-interaction resolution in the bars) and on the neutron-gamma separation (usually quantified in a Figure of Merit – FOM) in the energy range of interest. In practical applications outside of laboratories the ruggedness and stability of detectors against environmental factors as temperature variations or mechanical stress is also a major factor. Here, PU-based scintillators have a clear advantage over OGS [18]. This was our motivation to continue exploring PU-based scintillator recipes, to improve their performance, and to finally tap their full potential.

---

[1] Target Systemelektronik GmbH & Co. KG, Heinz-Fangman-Str. 4, 42287 Wuppertal, Germany, recently merged with Rapiscan Systems GmbH, Frankfurt.

This paper demonstrates the progress made with PU-based scintillators. It compares the latest PU-based scintillator material named M700 with the older version M600, and with the common commercial materials EJ-200 (for general use) and EJ-276D (providing PSD and an improved stability). Comparative measurements of light yield and PSD performance have been made before and after the scintillator samples were arbitrarily aged by "steaming" in a hot and humid atmosphere, followed by freezing in a refrigerator. This procedure was chosen to include the aspect of ruggedness in the evaluation, and to evidence the robustness of PU-based materials.

## 2. Materials and methods

*2.1. Production of the PU-based scintillator samples*

Scintillator samples of the PU-based materials M600 and M700 were produced inhouse at Rapiscan Systems' (former Target Systemelektronik's) production site in Wuppertal.

For the preparation of M600, 30 wt% PPO (2,5-diphenyloxazole) and 0.2 wt% Bis-MSB (1,4-bis(2-methylstyryl)benzene) – each relative to the total mass – were completely dissolved in the resin component. The curing agent was then added, and the reaction mixture was homogenized under stirring. The resulting formulation was transferred into a silicone mold, degassed under vacuum, and subsequently cured at 100 °C for 72 h.

The preparation of M700 followed an analogous procedure. In this case, however, 30 wt% m-terphenyl and 0.2 wt% E404 [2,2′-bis(7-(2,2-dimethylpropyl)-9,9-dipropyl-9H-fluoren-2-yl)-1,4-benzene] were used instead of PPO and Bis-MSB. All subsequent steps – including dissolution in the resin component, addition of the curing agent, homogenization, casting into a silicone mold, vacuum degassing, and curing at 100 °C for 72 h – were identical to the procedure described above.

*2.2. Scintillator materials investigated*

In addition to the M600 and M700 scintillators prepared inhouse, two ⌀26 mm × 26 mm-sized samples of common plastic scintillator materials – namely EJ-200 (general use, no PSD) and EJ-276D (providing PSD capabilities) – were purchased from Eljen Technology[2]. The general properties of all materials involved in this investigation are compiled in Table 1.

*2.3. Preparation and ageing of scintillator samples for the comparative measurements*

All scintillators, including the M600 and M700 samples prepared as described in 2.1, were cut to 10 mm height. The scintillator faces were sanded in several stages to at least 1500 grit, and finally polished to high gloss surface using a soft cloth disc. This in-house preparation warranted the same size and surface quality of all samples (Fig. 1a).

After performing an initial set of characterization measurements with the fresh samples, the scintillators were arbitrarily aged. First, they were packed in a temperature-controlled electrical cooking pot that was partially filled with water. A glass stand prevented the samples from directly touching the water level. The samples were "steamed" for 48 hours at 60°C with the lid closed, which ensures a relative humidity around 100 %. The ageing was continued by exposing all samples for 48 hours to – 18°C in a refrigerator. Finally, photographs were taken (Figure 1b). The EJ-200 and EJ-276D samples appeared clouded and milky, while the M600 and M700 looked undisturbed and transparent as before.

This procedure is much shorter and less intense than the accelerated ageing process proposed and used in [20]. However, the visible differences in the scintillators' appearance encouraged us to test their performance by another set of characterization measurements, with the intention to find differences in the scintillators' response to environmental effects.

---

[2] Eljen Technology, 1300 W. Broadway, Sweetwater, TX 79556. https://eljentechnology.com

*2.4. Characterization measurements*

Relative light yield, energy resolution, and pulse-shape discrimination performance of the scintillator samples were determined at a test stand in Rapiscan Systems' office in Wuppertal (Figure 2). The samples to be tested were not wrapped in any reflector but just fixed with optical grease (SilikonPaste 1001 Professional by NASP Lubricants) in the center of the light-sensitive face of a photomultiplier tube (PMT). The tube, a Hamamatsu R10601-100 with a photocathode of 34 mm diameter, was standing in upright position. The PMT head with the sample attached (Figure 2a) was first covered with a solid Teflon-coated cup (serving as reflector for the light leaving the scintillator) and a lightproof outer skin, then fitted with a tailored mu-metal shield and mounted in a protecting aluminum tube (Figure 2b), and finally concealed with a flexible mesh reducing high-frequency electromagnetic interferences (visible in Figure 2b). The voltage divider board was designed to provide very stable dynode voltages even at high PMT gain and fast, large current pulses. PMT output signals were tapped at the anode and the last dynode and fed AC-coupled (CR filter with 3 µs decay time) into two separate oscilloscope inputs. This provided separate trigger- and energy-signal channels. The input filters were set at 300 Mhz.

The scintillator was then simultaneously exposed to neutrons from an unmoderated $^{252}$Cf source emitting about 70,000 neutrons per second, and to gamma rays from a 20 kBq $^{137}$Cs source providing the reference spectrum for energy calibration. Gamma rays from the $^{252}$Cf source were shielded with a lead block of 20 mm thickness. The source-to-sample distances were adjusted to 10 cm. The dominating share of signals was due to $^{137}$Cs gamma rays. During exposure, the PMT anode current signal was sampled with a Teledyne LeCroy HDO6054-MS digital oscilloscope, triggered by the last-dynode signal, and stored on disk. The sampling rate was set to 2.5 GS/s; one pulse record comprised 2,500 samples or 1,000 ns at 12 bit resolution with a pre-trigger range of about 200 ns. Four hours of data collection, providing around 500,000 valid current pulses, were sufficient to characterize a single sample for energy depositions up to 1 MeVee (MeV electron equivalent).

For all measurements, the PMT position and orientation in the lab, the PMT voltage, and the oscilloscope settings were kept constant.

*2.5. Data analysis*

The data sets were analyzed offline with a dedicated Matlab script.

A pulse processing algorithm, applied to every signal record, realizes

1. The computation of a reference time (sample number) with a digital constant-fraction discriminator (CFD), and the derivation of "time zero" (with safe distance from the pulse onset) based on this reference time;

2. The baseline correction using a defined range of pre-trigger samples of the corresponding pulse record;

3. The integration of the baseline-corrected signal samples in a short (38-58 ns, depending on the material) and in a long (692 ns) integration gate starting at time zero, resulting in values $S_{short}$ and $S_{tot}$, respectively;

4. The computation of the tail-to-full ratio $TFR = \frac{S_{tail}}{S_{tot}} = \frac{S_{tot}-S_{short}}{S_{tot}}$, representing the share of the tail charge in the full charge of an anode current pulse. This is a common parameter often used for PSD.

The digital CFD is implemented as a Finite Impulse Response (FIR) filter followed by a zero-crossing detector, realizing a 4-bin (1.6 ns) smoothing, a 12-bin (4.8 ns) delay, and a CFD fraction of 0.5.

To translate $S_{tot}$ in a corresponding energy deposition, an $S_{tot}$ histogram is generated from all valid signals. As most signals are due to [137]Cs decays, this histogram exhibits a pronounced Compton edge at 477 keV, which represents the maximum energy deposition of Compton electrons due to scattered 662 keV gamma rays.

The Compton edge in a measured pulse-height spectrum is not an ideal step function but the result of (i) the underlying Compton electron energy distribution, (ii) multiple scattering and escape processes, and (iii) convolution with the detector response function. The resulting structure exhibits a characteristic asymmetric curvature with an inflection, and a single reversal point in its first derivative. An established method, proposed by Dietze and Klein [19], is fitting the experimental Compton edge against Monte-Carlo simulations convolved with a resolution function. This is physically well motivated, but it requires an accurate modeling of detector geometry and material composition, source location and geometry, assumptions about the energy-dependent resolution function, and reliable normalization and background treatment. Mismatches in these inputs can introduce systematic biases comparable to, or larger than, the statistical uncertainty of the edge determination. However, in the present study all materials are measured in identical geometries and under identical readout conditions. Light yields and energy resolutions of the materials investigated are quite similar. Moreover, the analysis focuses on relative light yields. Under these conditions, a low-order rational function provides a flexible yet well-constrained parametric representation of the Compton edge's shape: It captures the necessary curvature and asymmetry without over-parameterization, while remaining numerically stable under noise and limited statistics. In extensive internal testing, we found that the three-zero/two-pole form represents the local edge region robustly across different scintillator materials, count rates, and acquisition conditions, while avoiding the strong parameter correlations commonly observed in higher-order polynomial or spline fits. The reversal point of this rational function is uniquely defined and reproducible, making it well suited as an operational definition of the Compton-edge reference point.

Therefore, the edge is fitted with a rational function comprising three zeros and two poles, and the reversal point of this function is taken as the reference point $S_{tot}^{477}$ corresponding to the Compton-edge energy (Figure 3). The pulse processing with baseline subtraction yields $S_{tot} = 0$ for zero energy deposition; therefore, the $S_{tot}$ scale can be translated in an energy scale by

$$E = c \cdot S_{tot} \text{ with } c = \frac{477 \text{ keVee}}{S_{tot}^{477}},$$

assuming a linear relation between both values. Note that this scale, labeled in keV electron equivalent (keVee), is valid just for electrons interacting with the scintillator; a translation to recoil proton energies must consider the light quenching [1], [17].

Furthermore, the Compton edge position $S_{tot}^{477}$ measures the relative light yield of a sample. In our analysis, the light yield of EJ-200 before ageing was taken as reference. The ratio

$$LY_{rel} = S_{tot}^{477}(\text{X})/S_{tot}^{477}(\text{EJ-200})$$

then represents relative light yield of another sample X. Note that the relative light yield determined in this way relates to PMT readout with the spectral sensitivity of the given type (Hamamatsu R10601-100); spectral effects have not been considered.

After calibration, an energy spectrum (actually, the distribution of energy depositions $E$ in keVee) and a PSD plot (two-dimensional distribution of $TFR$ versus $E$) are generated comprising all triggered events of a data set. The projection of multiple (thin) energy slices on the $TFR$ axis results in $TFR$ distributions with two peaks corresponding to neutrons (recoil protons) or gamma rays (recoil electrons). A Figure of Merit

$$FOM = \frac{P_n - P_\gamma}{FWHM_n + FWHM_\gamma}$$

with $P_n$ and $P_\gamma$ meaning the positions, $FWHM_n$ and $FWHM_\gamma$ the full widths at half maximum of the neutron and gamma peaks determined by fitting two Gaussian functions, respectively, quantifies the PSD performance of a scintillator sample at the given energy deposition.

As average pulse shapes are expected to vary with the scintillator material (which is the reason for varying PSD performance), the short integration gate was separately optimized for each of the materials investigated. For this purpose, the gate width was varied between 30 and 60 ns in steps of 5 pulse samples (2 ns), and the gate length providing the maximum FOM for energy depositions between 452 and 502 keVee (i.e., an energy slice around the 477 keV Compton edge of $^{137}$Cs gamma rays) was taken for generating the figures given in this paper.

PSD plots, $TFR$ distributions in an energy window comprising the 477 keV reference Compton edge, and FOM dependencies on the energy deposition derived from slicing the PSD plot in the full energy range of interest and fitting the resulting FOMs with a rational function comprising three zeros and two poles, are shown in Figure 4.

*2.6. Uncertainties*

To estimate the uncertainty of relative light yields obtained with this procedure, a reproducibility test was performed after the initial characterization measurements. The EJ-200 sample was six times mounted on the PMT as described in section 2.4, measured with the respective procedure and the same statistics as in the characterization runs, and then removed from the PMT. The tests were performed on different days during a week. The Compton edge position was determined as described above. The statistical uncertainties of the Compton-edge positions, derived from uncertainties of the fit parameters provided by the fit procedure, was in the range of 1.7 to 2.2 % with an average of 1.97 %. The standard deviation of the Compton-edge positions obtained in the six runs was only 0.74 %, i.e., smaller than the statistical uncertainty estimate. Therefore, we assume an uncertainty of 2.0 % for the relative light yield determination.

The good reproducibility is obviously due to the following facts:

(i) The samples were not wrapped but positioned in a reflective cup. This eliminates light yield variations due to imperfect wrapping.

(ii) Temperature variations in the lab were quite small and well below ± 2°C. Temperature effects on the PMT gain are thus negligible.

(iii) The PMT orientation did not change because of the base fixed in the setup. This and the magnetic shield reduced potential effects of the terrestrial magnetic field.

Uncertainties of the FOMs are derived by error propagation from uncertainties of the peak positions and peak widths obtained from the corresponding fit procedure. Error bands of the FOM fits are provided by the fit algorithm.

## 3. Results and discussion

The results of this work are presented in Figure 4 to Figure 7 and in Table 2.

Figure 4 exhibits the quite decent neutron-gamma discrimination capabilities of EJ-276D, M600, and M700 scintillators in contrast to the rudimentary (but, surprisingly, still observable) PSD performance of the common EJ-200. The data were taken before ageing. It is evident that M700 provides the best neutron-gamma discrimination, reaching a FOM > 2 below 200 keVee and keeping a good discrimination (FOM > 1.5) down to 100 keVee (Figure 6). The performance of M600 is worse than that of EJ-276D; the differences seem to be larger at lower energies (Figure 6), which might be due to the lower light yield (Table 2). Altogether, all three scintillators offered with PSD capabilities are good candidates for simultaneous measurements of fast neutrons and gamma rays.

The characterization measurements were repeated after ageing of all samples by steaming and freezing as described in section 2.3. Though this procedure was rather short if compared with other, similar investigations [20], an optical degradation of the Eljen materials EJ-200 and EJ-276D was clearly visible: The corresponding samples became milky and opaque, while the PU-based scintillators, M600 and M700, seemed undisturbed and clear as before (see Figure 1). Surprisingly, these visible differences are not reflected in the test results, at least not to the extent expected. Light yield and FOM measured at 477 keVee (Table 2) show only slight variations which do not significantly exceed the experimental uncertainties. Only the overall FOM trend (Figure 5) indicates a minor degradation above 500 keVee for the EJ-276D and M700 materials, while M600 seems unaffected.

Figure 6 summarizes the discrimination performance of the three PSD scintillators before and after ageing in terms of their FOM plotted versus energy deposition. M700 provides the best separation, followed by EJ-276D and M600. This picture does not change with the ageing as performed in this work.

Characteristic pulse shapes of neutron- and gamma-induced signals in EJ-276D and M700 are given in Figure 7. The graphs represent sums of all pulses with energy depositions between 452 and 502 keVee which are identified as neutron or gamma-ray responses, respectively, with the pulse maximum normalized to 1. For comparison, a normalized sum of single-electron pulses (SEP) is also shown, together with a SEP model representing a delta function (the initial photoelectron emission), convolved with a Gaussian ($\sigma = 7$ ns) representing the PMT response to a single photoelectron, the $\tau = 9.5$ ns exponential decay of the fast RC filter at the oscilloscope input (300 MHz bandwidth limitation), and the $\tau = 3$ $\mu$s differentiated exponential due to the AC-coupled anode readout. This model describes the SEP response very well. On the one hand it is obvious that the leading part of all scintillator signals is strongly affected, in fact rather determined, by this SEP response. On the other hand, the pulse tail is indeed very different for the gamma- and neutron induced signals, but also for the distinct scintillator materials. The M700 signals are distinguished by a shorter decay, which generally allows a shorter full-charge integration time. The latter is of importance for applications where high detector loads must be handled, as shorter integration gates reduce the dead time per event and increase the maximum throughput of the pulse-processing electronics.

The light yields relative to the EJ-200 sample, and the FOM values, both determined for the 477 keV Compton edge before and after the ageing process, are compiled in Table 2.

The light yields of EJ-200, EJ-276D, and M700 are very similar, and they are only marginally affected by the "steaming" and "freezing" processes. The light yield of M600 is generally lower. While M600 also shows a smaller FOM, reflecting a lower PSD performance, the M700 can compete with EJ-276D in all respects.

The relatively short ageing procedure applied here is obviously not sufficient to disclose differences in the long-term stability of various materials. Likewise, this study has not touched properties as stress behavior, breaking resistance, elasticity, plasticity – properties that determine the care and expense needed for mechanical processing. Further investigations should also include self-absorption effects which are important for applications in RPMs or large-area neutron imagers, as they degrade the performance of large-size scintillators. Altogether it would be very useful to include PU-based scintillators in future studies targeting the usability, ruggedness, and long-term stability of organic scintillators in various applications.

## 4. Applications of PU-based scintillators

PU-based scintillators, especially the latest version M700, are suitable materials for neutron or dual-particle (neutron and gamma-ray) imagers as they combine competitive PSD performance with mechanical robustness and ruggedness. Bar-shaped M600 and M700 scintillators fitting with the design of NOVO[3], a modular detector array for simultaneous neutron and gamma-ray imaging in cancer treatments with proton beams [16], have already been produced [18] and will be tested soon under realistic treatment conditions in a proton therapy center. A first imaging experiment was already performed at an accelerator-based neutron source, using a provisional setup comprising OGS and M600 scintillator bars of the same size ($12 \times 12 \times 140$ mm$^3$), read out on both sides with Hamamatsu S14161-3050HS-04 silicon photomultiplier (SiPM) arrays coupled to frontend electronics by Target Systemelektronik [21]. The results confirmed the detector concept as well as the usability of both scintillator materials, OGS and M600, for the NOVO project.

Another obvious application is radiation protection and dosimetry. The oxygen and nitrogen content of PU-based scintillators makes their response more body-equivalent than that of classic plastic scintillators, meaning their integrated signal (the total light produced in the scintillators) in reference gamma radiation fields better approaches the energy dependence of the dose absorbed by a human body. This suggests using this material for dose monitoring. Corresponding explorations are in progress.

Generally, PU-based scintillators could replace other plastics if the higher density and ruggedness are of importance.

---

[3] https://www.novo-project.eu

## 5. Conclusions

This paper compares small samples of two polyurethane-based scintillator materials – the older M600 and the recent M700 by Rapiscan Systems – with Eljen EJ-276D, an up-to-date commercial plastic scintillator with neutron-gamma discrimination capabilities. M700 provides the best PSD performance and about the same light yield as EJ-276D, while its light pulses exhibit a shorter pulse decay. This qualifies the material for high-rate applications. The accelerated ageing process applied in this work was too short to trigger distinct performance degradation in any of the materials, though optical degradation was already visible. Nevertheless, the extremely robust polyurethane matrix promises advantages in resilience and long-term performance. PU-based scintillators should thus be included in future usability studies of organic scintillators in various applications.

**CRediT authorship contribution statement**

Olga Maiatska: Conceptualization, methodology - chemistry, sample preparation (lead), writing – original draft preparation, review and editing

Torsten Dünnebacke: Methodology – chemistry, sample preparation, writing – review and editing

Martin Kreuels: Methodology – sample preparation and finishing, writing – review and editing

Guntram Pausch: Conceptualization, writing – original draft preparation (lead), review and editing

Falko Scherwinski: Methodology – measuring setup, writing – review and editing

Jürgen R. Stein: Supervision, conceptualization (lead), software, formal analysis, visualization, writing – original draft preparation, review and editing

**Declaration of generative AI and AI-assisted technologies in the writing process**

No generative AI or AI-assisted technologies have been used in the writing process of this publication.

**Declaration of competing interest**

All authors are employed by Rapiscan Systems GmbH, the owner of the M600 / M700 production technologies.

**Data availibility**

Data will be made available on request.

Table 1.

Basic properties of the scintillator materials investigated.

| Scintillator | EJ-200 [a] | EJ-276D [a] | M600 | M700 |
|---|---|---|---|---|
| Polymer base | Polyvinyl Toluene (PVT) | | Polyurethane (PU) | |
| Density (g/cm$^3$) | 1.023 | 1.099 | 1.207 | 1.164 |
| Atoms per cm$^3$ ($\times 10^{22}$) | | | | |
|   – Hydrogen | 5.17 | 4.65 | 5.69 | 4.53 |
|   – Carbon | 4.69 | 4.94 | 3.96 | 4.24 |
|   – Nitrogen | 0.0 | 0.0 | 0.40 | 0.35 |
|   – Oxygen | 0.0 | 0.0 | 0.85 | 0.59 |
| Softening point | 75°C | [b] | 72°C | 75°C |
| Shore "D" hardness | 84 | [b] | 86 | 82 |
| Refractive index | 1.58 | [b] | 1.57 | 1.62 |
| Wavelength of maximum emission (nm) | 425 | 425 | 410 | 400 |

[a] data from Eljen website https://eljentechnology.com/products/plastic-scintillators
[b] not specified

Table 2.

Optimized short-gate lengths ($t_{short}$), relative light yield ($LY_{rel}$) and Figure of Merit (FOM) measured before and after ageing.

| Scintillator | | EJ-200 | EJ-276D | M600 | M700 |
|---|---|---|---|---|---|
| $t_{short}$ | optimized for best PSD | 38 ns | 58 ns | 42 ns | 46 ns |
| $LY_{rel}$ [a] | before ageing | 100.0 % [c] | (98.1 ± 2.0) % | (81.3 ± 2.0) % | (99.9 ± 2.0) % |
| | after ageing | (98.0 ± 2.0) % | (98.6 ± 2.0) % | (79.7 ± 2.0) % | (99.1 ± 2.0) % |
| | LY change | (-2.0 ± 2.0) % | (0.5 ± 2.8) % | (-2.0 ± 2.8) % | (-0.8 ± 2.8) % |
| FOM [b] | before ageing | 0.40 ± 0.02 | 2.26 ± 0.03 | 1.95 ± 0.03 | 2.83 ± 0.04 |
| | after ageing | 0.34 ± 0.02 | 2.31 ± 0.03 | 1.93 ± 0.03 | 2.76 ± 0.04 |
| | FOM change | (-15.0 ± 7.1) % | (2.2 ± 1.9) % | (-1.0 ± 2.2) % | (-2.1 ± 2.0) % |

[a] relative to EJ-200 before ageing
[b] energy range 452-502 keVee
[c] reference value of the relative measurement

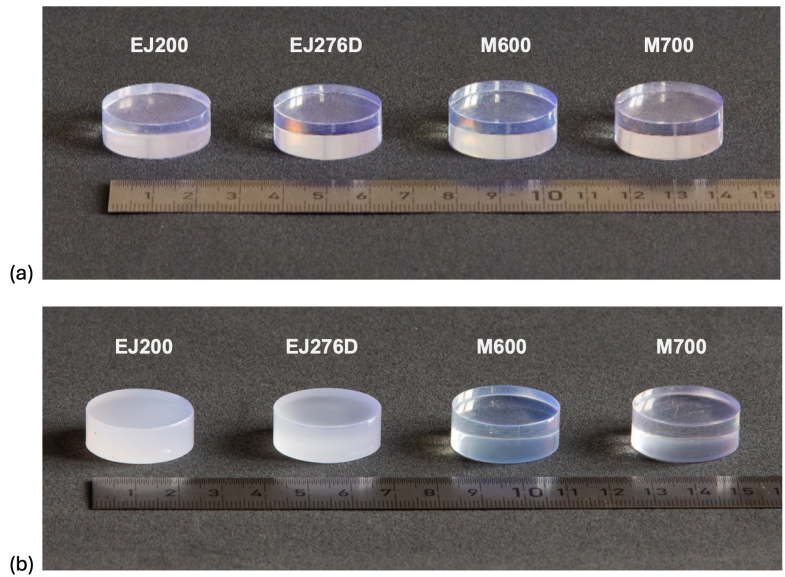

Figure 1. Photographs of the scintillator samples (a) after sample preparation, (b) after 48 hours of "steaming" at 60°C and 48 hours of "freezing" at –18°C. The EJ-200 and EJ-276D samples clearly show degraded optical quality (milky appearance), while the PU-based M600 and M700 samples keep their initial transparency.

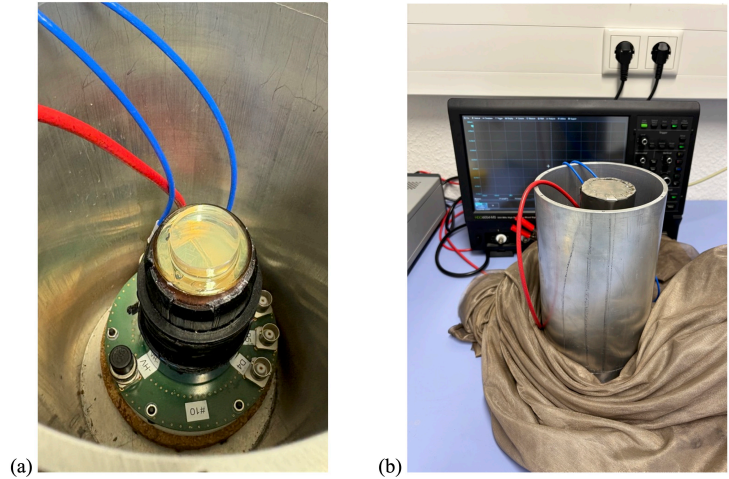

(a)　　　　　　　　　　　　　　(b)

Figure 2.　Photographs of the setup for scintillator characterization: (a) Photomultiplier tube (PMT) standing on the voltage divider, with a scintillator sample fixed on top. The red cable feeds high voltage to the HV divider, the blue cables are used for signal extraction. (b) PMT with the scintillator sample covered by a Teflon-coated cap, fitted with a mu-metal shield, mounted in a protecting tube. A flexible, RF-shielding fine metal mesh, here wrapped around the protecting tube, is finally thrown over the setup.

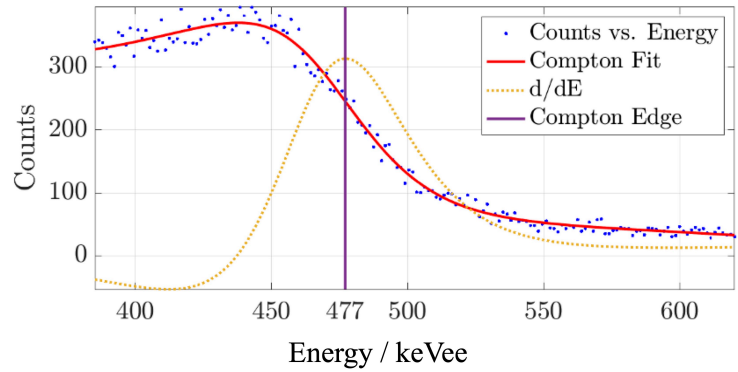

Figure 3. Illustration of the energy calibration. The histogram of long-gate integrals $S_{tot}$ reflecting the total pulse charges of a complete data set exhibits a pronounced Compton edge due to 662 keV gamma rays from the $^{137}$Cs source. The edge is fitted with a rational function, and the reversal point of this function is taken as reference point $S_{tot}^{477}$ corresponding to the Compton-edge energy, which allows translating the $S_{tot}$ numbers in electron-equivalent energy deposition.

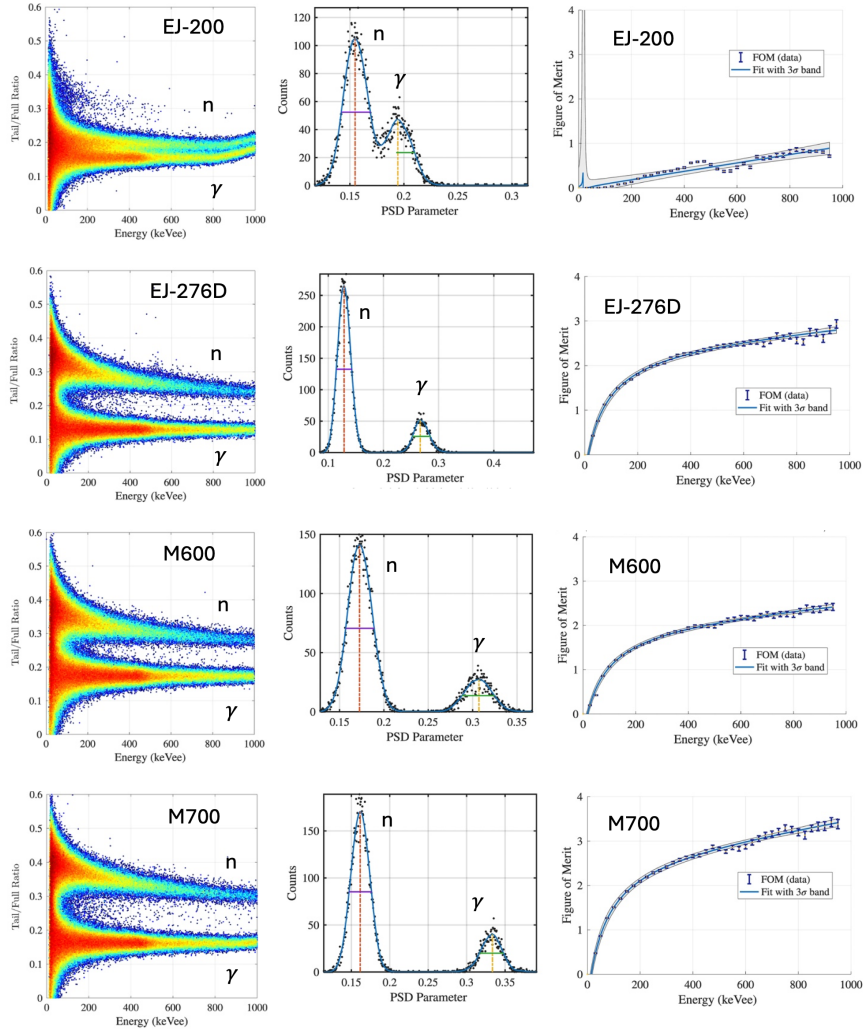

Figure 4. PSD performance of the scintillators investigated (before ageing). The left panels represent *TFR-E* distributions. The *TFR* spectra (middle) comprise events of the energy bin 452 to 502 keVee; they visualize the neutron-gamma separation of all PSD scintillators around the 477 keV Compton edge, and the fits used to determine a Figure of Merit (FOM). Corresponding fits performed for multiple energy bins yield the energy dependence of the FOM, shown in the right panels. The FOM data points are then fitted with a rational function.

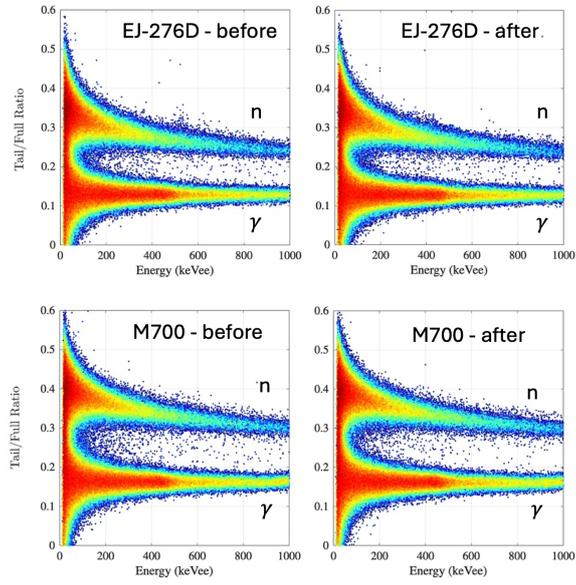

Figure 5. Comparison of the PSD performance of EJ-276D and M700 before and after ageing.

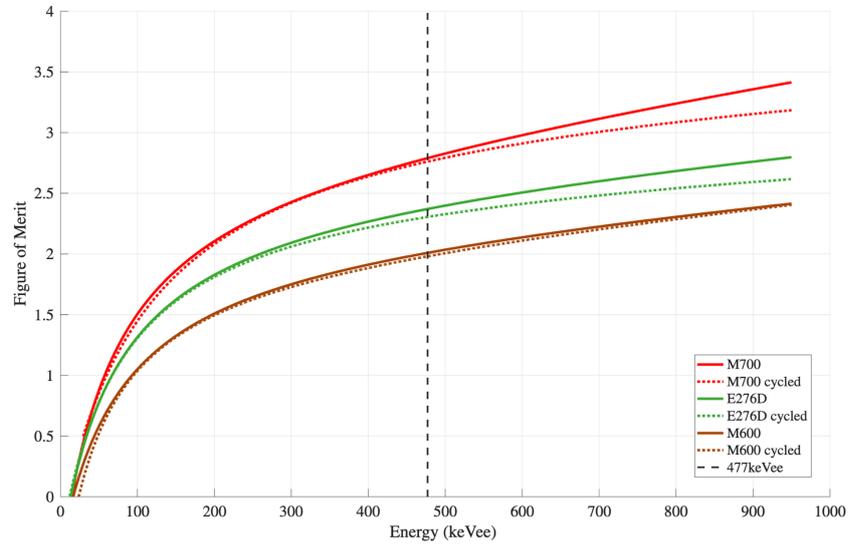

Figure 6.	Dependence of the PSD Figure-of-Merit (FOM) on the energy deposition for the PSD-capable scintillators investigated. Solid lines – before ageing; dotted lines – after ageing.

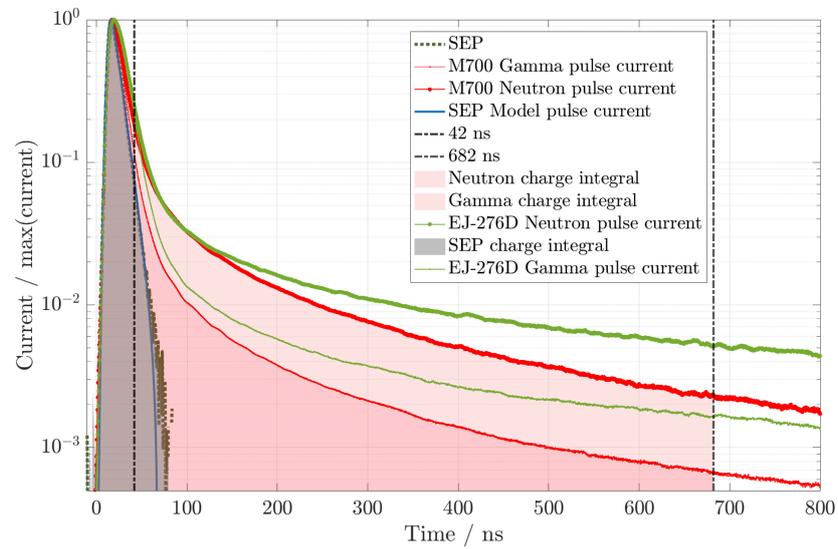

Figure 7. Comparison of EJ-276D and M700 pulse shapes. The single-photoelectron pulse (SEP) shape is also shown together with a corresponding model (see text). Note the shorter pulse decay of the M700 signals which allows shorter full-charge integration gates.